\documentclass[twocolumn]{aastex63}

\usepackage{natbib}
\usepackage[style=american]{csquotes}
\usepackage{amsmath}
\usepackage{float}
\usepackage[caption = false]{subfig}
\newcommand{\tess}{\emph{TESS}}

\providecommand{\bjdtdb}{\ensuremath{\rm {BJD_{TDB}}}}

\providecommand{\mj}{\ensuremath{\,M_{\rm J}}}
\providecommand{\rj}{\ensuremath{\,R_{\rm J}}}

\providecommand{\fluxcgs}{10$^9$ erg s$^{-1}$ cm$^{-2}$}

\providecommand{\logg}{cm\,s$^{-2}$}

\shorttitle{TOI-1994\,b}
\shortauthors{Page, et al.}

\begin{document}

\title{TOI-1994\,b: A Low Mass Eccentric Brown Dwarf Transiting A Subgiant Star}

\correspondingauthor{Emma Page}
\email{ejp520@lehigh.edu}

\author[0000-0002-3221-3874]{Emma Page}
\affiliation{Department of Physics, Lehigh University, 16 Memorial Drive East, Bethlehem, PA 18015, USA}

\author[0000-0002-3827-8417]{Joshua Pepper}
\affiliation{Department of Physics, Lehigh University, 16 Memorial Drive East, Bethlehem, PA 18015, USA}

\author[0000-0001-7294-5386]{Duncan Wright}
\affiliation{University of Southern Queensland, Centre for Astrophysics, USQ Toowoomba, West Street, QLD 4350 Australia}

\author[0000-0001-8812-0565]{Joseph E. Rodriguez}
\affiliation{Center for Data Intensive and Time Domain Astronomy, Department of Physics and Astronomy, Michigan State University, East Lansing, MI 48824, USA}

\author[0000-0001-9957-9304]{Robert A. Wittenmyer}
\affiliation{University of Southern Queensland, Centre for Astrophysics, USQ Toowoomba, West Street, QLD 4350 Australia}

\author[0000-0002-7084-0529]{Stephen R. Kane}
\affiliation{Department of Earth and Planetary Sciences, University of California, Riverside, CA 92521, USA}

\author[0000-0003-3216-0626]{Brett Addison}
\affiliation{University of Southern Queensland, Centre for Astrophysics, USQ Toowoomba, West Street, QLD 4350 Australia}
\affiliation{Swinburne University of Technology, Centre for Astrophysics and Supercomputing, John Street, Hawthorn, VIC 3122, Australia}


\author{Timothy Bedding}
\affil{School of Physics, Sydney Institute for Astronomy (SIfA), The University of Sydney, NSW 2006, Australia}

\author{Brendan P. Bowler}
\affil{Department of Astronomy, The University of Texas at Austin, TX 78712, USA}

\author[0000-0001-7139-2724]{Thomas~Barclay}
\affiliation{NASA Goddard Space Flight Center, 8800 Greenbelt Road, Greenbelt, MD 20771, USA}
\affiliation{University of Maryland, Baltimore County, 1000 Hilltop Circle, Baltimore, MD 21250, USA}

\author[0000-0001-6588-9574]{Karen A.\ Collins}
\affiliation{Center for Astrophysics \textbar \ Harvard \& Smithsonian, 60 Garden Street, Cambridge, MA 02138, USA}


\author[0000-0002-5674-2404]{Phil Evans}
\affiliation{El Sauce Observatory, Coquimbo Province, Chile}

\author[0000-0002-1160-7970]{Jonathan Horner}
\affil{University of Southern Queensland, Centre for Astrophysics, West Street, Toowoomba, QLD 4350 Australia}

\author[0000-0002-4625-7333]{Eric L.\ N.\ Jensen}
\affiliation{Department of Physics \& Astronomy, Swarthmore College, Swarthmore PA 19081, USA}

\author[0000-0002-5099-8185]{Marshall C. Johnson}
\affiliation{Department of Astronomy, The Ohio State University, 4055 McPherson Laboratory, 140 West 18$^{\mathrm{th}}$ Ave., Columbus, OH 43210 USA}

\author{John Kielkopf}
\affil{Department of Physics and Astronomy, University of Louisville, Louisville, KY 40292, USA}



\author[0000-0002-4510-2268]{Ismael~Mireles}
\affiliation{Department of Physics and Astronomy, University of New Mexico, 210 Yale Blvd NE, Albuquerque, NM 87106, USA}


\author{Peter Plavchan}
\affil{George Mason University, 4400 University Drive MS 3F3, Fairfax, VA 22030, USA}

\author[0000-0002-8964-8377]{Samuel N. Quinn}
\affiliation{Center for Astrophysics \textbar \ Harvard \& Smithsonian, 60 Garden Street, Cambridge, MA 02138, USA}

\author[0000-0002-6892-6948]{S.~Seager}
\affiliation{Department of Physics and Kavli Institute for Astrophysics and Space Research, Massachusetts Institute of Technology, Cambridge, MA 02139, USA}
\affiliation{Department of Earth, Atmospheric and Planetary Sciences, Massachusetts Institute of Technology, Cambridge, MA 02139, USA}
\affiliation{Department of Aeronautics and Astronautics, MIT, 77 Massachusetts Avenue, Cambridge, MA 02139, USA}


\author[0000-0002-3481-9052]{Keivan G.\ Stassun}
\affiliation{Department of Physics and Astronomy, Vanderbilt University, Nashville, TN 37235, USA}

\author[0009-0008-5145-0446]{Stephanie Striegel}
\affiliation{SETI Institute, Mountain View CA 94043/NASA Ames Research Center, Moffett Field CA 94035}   


\author[0000-0002-4265-047X]{Joshua N.\ Winn}
\affiliation{Department of Astrophysical Sciences, Princeton University, Princeton, NJ 08544, USA}

\author[0000-0002-4891-3517]{George Zhou}
\affiliation{University of Southern Queensland, Centre for Astrophysics, USQ Toowoomba, West Street, QLD 4350 Australia}


\author{Carl Ziegler}
\affiliation{Department of Physics, Engineering and Astronomy, Stephen F. Austin State University, 1936 North St, Nacogdoches, TX 75962, USA}

\begin{abstract}
We present the discovery of TOI-1994\,b, a low-mass brown dwarf transiting a hot subgiant star on a moderately eccentric orbit. TOI-1994 has an effective temperature of $7700^{+720}_{-410}$\,K, V magnitude of 10.51\,mag and log(g) of $3.982^{+0.067}_{-0.065}$. The brown dwarf has a mass of $22.1^{+2.6}_{-2.5}$\mj, a period of 4.034 days, an eccentricity of $0.341^{+0.054}_{-0.059}$, and a radius of $1.220^{+0.082}_{-0.071}$\rj. TOI-1994\,b is more eccentric than other transiting brown dwarfs with similar masses and periods.
The population of low mass brown dwarfs may have properties 
similar to planetary systems if they were formed in the same way, but the short orbital period and high eccentricity of TOI-1994\,b may contrast this theory. An evolved host provides a valuable opportunity to understand the influence stellar evolution has on the substellar companion's fundamental properties. With precise age, mass, and radius, the global analysis and characterization of TOI-1994\,b augments the small number of transiting brown dwarfs and allows the testing of substellar evolution models.  
\end{abstract}


\section{Introduction} \label{intro}
Brown dwarfs are substellar objects that burn deuterium and have masses between those of planets and stars. A compact object's ability to burn deuterium is primarily defined by the mass of the object, but the most common definition is debated in the astronomical community. The International Astronomical Union (IAU) uses the limiting mass for thermonuclear fusion of deuterium as the lower limit for brown dwarfs, which is around $13$ $ M_{J}$ for objects of solar metallicity \citep{Boss}.

There are arguments that the definition of a brown dwarf should instead be based on formation mechanisms rather than mass. They suggest that large planets can burn deuterium similar to brown dwarfs, and that the two domains overlap when using the IAU definition. This new definition based on formation mechanisms of brown dwarfs considers whether the substellar object possibly formed more like a star or like a planet. This suggests that the brown dwarf regime is simply the tail end of the planetary and stellar formation mechanisms. Solidifying this definition proves to be difficult since formation theory is incomplete, the boundary between formation mechanisms is unclear, and we have no definitive way of knowing how a particular object formed \citep{RevModPhys.73.719, 2021A&A...652A.127G}.    

It is possible that different formation mechanisms lead to different populations of brown dwarf companions. \citet{10.1093/mnras/stu134} discovered that the brown dwarf population can be split at around 42.5 $M_J$ based on eccentricity. Lower mass brown dwarfs display an eccentricity distribution consistent with massive planets, while higher mass brown dwarfs have an eccentricity distribution similar to binary stars. This could suggest that brown dwarfs with mass $<42.5 M_J$ form similar to planets in protoplanetary disks and $>42.5 M_J$ form similarly to stellar binaries (although see \citet{Schlaufman2018} for an alternate proposal to this boundary).  More recent studies found a similar brown dwarf eccentricity distributions using larger populations \citep{10.1093/mnras/stx334, refId0, 2021A&A...652A.127G}.

Compared to giant planets and stars, brown dwarfs are less often detected as short period companions around main sequence stars \citep{2006ApJ...640.1051G, 2011A&A...525A..95S}. Although this result is slowly changing in recent years due to space based photometry, brown dwarfs still have a lower occurrence rate according to more recent studies (eg., \cite{10.1093/mnras/stx334, refId0}). The lack of brown dwarf companions with periods less than 100 days is commonly referred to as the ``brown dwarf desert.'' This desert may be caused by the different formation mechanisms for low mass and high mass brown dwarfs. The transition point between planet formation and binary star formation is not fully understood. Well characterized brown dwarf companions are required to provide insight into the brown dwarf desert and formation mechanisms. Specifically, transiting brown dwarf companions with precisely calculated radii are quite valuable. With a tight age, and precise brown dwarf mass, radius and eccentricity, we can directly test substellar evolution models (eg. \cite{2021AJ....161...97C, 2023arXiv230109663V}).

Space-based photometric missions make discovering transiting brown dwarfs more accessible. The Transiting Exoplanet Survey Satellite \citep[\tess;][]{TESS} is a space-based photometric survey with a primary mission of discovering transiting exoplanets. \tess{} was launched in 2018 and has currently confirmed 330 planets with over 6,400 planet candidates identified as of 24 April 2023 \footnote{\url{https://exoplanetarchive.ipac.caltech.edu/}}, as well as monitoring known exoplanetary systems \citep{Kane2021}.
In addition to planets, \tess{} is able to detect transiting brown dwarf companions.  Out of the 39 confirmed transiting brown dwarf companions currently known, \tess{} has discovered almost half \citep{2021A&A...652A.127G, 2022AJ....163...89C, 2022A&A...664A..94P,  2022MNRAS.514.4944C, 2022MNRAS.516..636S, 2022arXiv221013939L, 2023arXiv230109663V}. The use of \tess{} to detect transiting brown dwarf companions enables the determination of precise radii, and may therefore illuminate the brown dwarf desert and associated formation mechanisms.


We report the discovery of TOI-1994\,b, a low mass brown dwarf with moderate eccentricity transiting an evolved star. TOI-1994\,b is one of only six known brown dwarf companions transiting an evolved star \citep{2021AJ....161...97C, 2022A&A...664A..94P, 2022arXiv221013939L}. The moderate eccentricity of TOI-1994\,b may present a discrepancy with previous attempts to define brown dwarfs using eccentricity distributions to determine formation mechanisms. 

In Section \ref{sec:observations}, we present the TESS discovery and spectroscopic follow up observations for TOI-1994\,b. Section \ref{sec:exofast} describes the EXOFASTv2 global fit and presents final parameters for the TOI-1994 system. Section \ref{sec:discussion} discusses the ways that TOI-1994\,b is unique and how it compares to other brown dwarf companions. 

\section{Discovery and Follow Up Observations} \label{sec:observations}
\subsection{TESS Discovery} \label{sec:TESS}
\tess{} \footnote{All the {\it TESS} data used in this paper can be found in MAST: \dataset[10.17909/t9-nmc8-f686]{http://dx.doi.org/10.17909/t9-nmc8-f686} and \dataset[10.17909/t9-r086-e880]{http://dx.doi.org/10.17909/t9-r086-e880}.} \citep{TESS} is a near all-sky photometric survey searching for exoplanets transiting mainly nearby bright main sequence stars. \tess{} splits most of the sky into sectors and observes each for a minimum time baseline of 27 days. During the first two years of the mission, \tess{} observed about 160,000 selected stars at a 2-min cadence, and obtained 30-min cadence Full Frame Images (FFIs) for the entire sector region. In July of 2022, \tess{} completed its primary mission and began its first Extended Mission which switched to a 10-min cadence for FFIs. Starting in September of 2022, \tess{} began its second Extended Mission and increased the cadence for the FFIs to 200 seconds.

\tess{} observed TIC 445903569 in sectors 9, 10, and 36. \tess{} obtained 30-min FFIs in sectors 9 and 10 and 10-min FFIs in sector 36.  In addition, TIC 445903569 was observed during sector 36 in 2-min cadence. Initial transits were detected by the Quick-Look Pipeline \citep[QLP;][]{QLP} using data taken from the 30-min FFIs from sectors 9 and 10. In October of 2021, the \tess{} Object of Interest (TOI) team at MIT identified TIC 445903569 as a host star candidate using the QLP light curves, and designated it TOI-1994 using the process described in \citet{2021ApJS..254...39G}. The later 2-min cadence \tess{} observations from sector 36 were processed using the Science Processing Operations Center \citep[SPOC;][]{SPOC} pipeline from the NASA Ames Research Center, which produced PDC-SAP lightcurves we used \cite{Stumpe2012, Stumpe2014, Smith2012}. The SPOC conducted a transit search of Sector 36 on 14 April 2021 with an adaptive, noise-compensating matched filter \citep{jenkins2002, jenkins2010, jenkins2020} 
, producing a TCE for which an initial limb-darkened transit model was fitted \citep{Li2019} and a suite of diagnostic tests were conducted to help make or break the planetary nature of the signal \citep{Twicken2018}. The signal was repeatedly recovered as additional observations were made in sectors 62 and 63, and the transit signature passed all the diagnostic tests presented in the Data Validation reports. The host star is located within $2.09 \pm 2.51$ arcsec of the source of the transit signal. We used the \texttt{lightkurve} python package \citep{lightkurve} to collect the light curves from the Mikulski Archive for Space Telescopes (MAST) and flattened the data using \texttt{Keplerspline} \footnote{\url{https://github.com/avanderburg/keplerspline}}.  

\begin{figure*}
    \centering
    \includegraphics[scale=.5]{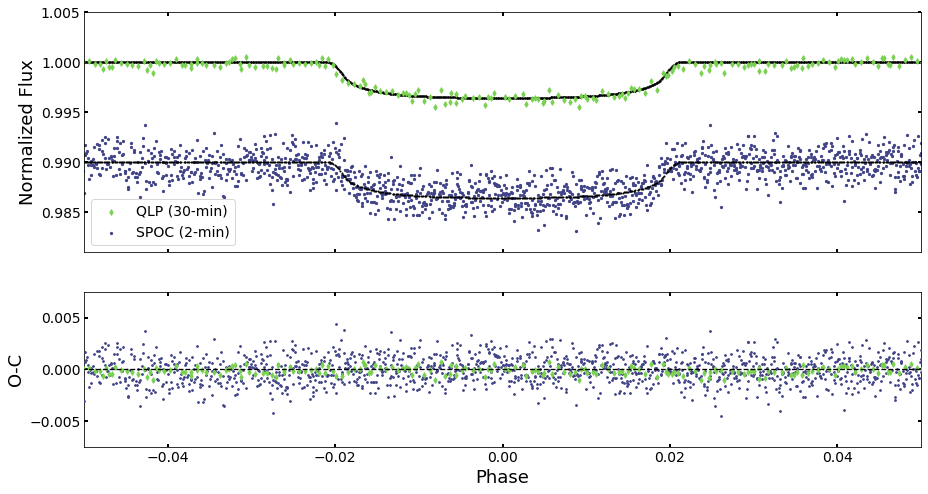}
    \caption{Discovery light curves from \tess{} for TOI-1994\,b using 2-min cadence data from the SPOC pipeline from Sector 36 (blue dots) and 30-min cadence data from the QLP pipeline from Sectors 9 and 10 (green diamonds). The black line represents a model fit from the EXOFASTv2 described in Section \ref{sec:exofast}. The curves are phase-folded to a period of $4.0337142$ days. The residuals for the best fit model are shown in the bottom panel.}
    \label{fig:TESS}
\end{figure*}

Figure \ref{fig:TESS} shows the light curves of TIC 445903569 from the SPOC and QLP pipelines, folded to the period of the transit. The transits have a depth of $3.23$ ppt, a period of $4.0337142$ days, and duration of $3.82$ hours. 

\subsection{Follow Up Photometry} \label{sec:phot}
We acquired ground-based time-series photometry as part of the TESS Follow-up Observing Program (TFOP; \cite{Collins2018}). We used the TESS Transit Finder, a customized version of the Tapir software package \citep{Jensen2013} to schedule our observations. 

We observed three transit events with the Las Cumbres Observatory Global Telescope Network \citep[LCOGT;][]{Brown13} 1 m telescopes and Sinistro cameras. We observed an egress on 2020 November 23 in the $z$ band from the Siding Spring Observatory node, a full transit on 2021 February 8 in $z$ from Cerro Tololo Interamerican Observatory (CTIO), and a full transit minus egress on 2021 February 15 in $B$, again from CTIO. The data were reduced using the LCOGT facility BANZAI pipeline \citep{McCully18}, and the light curves extracted using AstroImageJ \citep{Collins17}.
 
We observed a full transit in Johnson-Cousins $R_c$-band on UT20210207 using the Evans 0.36m telescope at El Sauce Observatory in Coquimbo Province, Chile. The telescope was equipped with a ST1603-3 CCD camera with 1536 x 1024 pixels binned 2 x 2 in-camera resulting in an image scale of 1.47"/pixel. The photometric data was obtained from 116 x 25 secs and 870 x 30 secs exposures, after standard calibration, using a circular 7.4" aperture in AstroImageJ \citep{Collins17}. 

These ground-based photometry showed that the transit occurred on-target on the expected host star, and was achromatic, thus helping to vet the candidate for further follow-up observations.

\begin{figure}
    \centering
    \includegraphics[width=.5\textwidth]{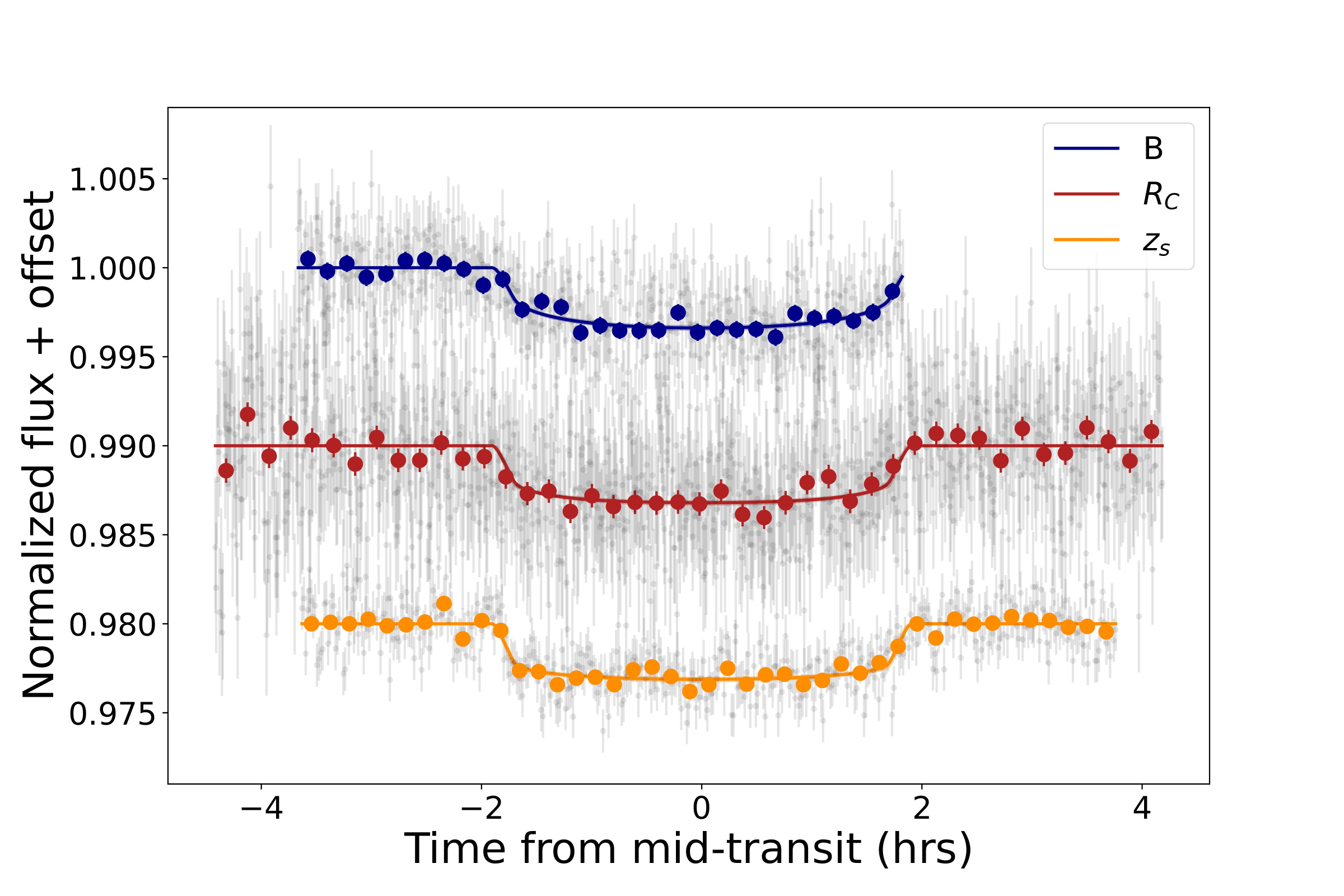}
    \caption{Ground-based lightcurves taken in B, Rc and PANSTARRS z-short passbands plotted with corresponding models. The depth of the transit is consistent across these passbands.   }
    \label{fig:followup}
\end{figure}

We jointly modeled all three ground-based light curves (taken in the $B$, $R_c$, and PANSTARRS $z$-short passbands) and allowed the dilution to vary independently for two of the bands relative to the third, mimicking the case where the occulting body or a third source contributes significantly to the observed lightcurve. Figure \ref{fig:followup} displays these three ground-based lightcurves and the joint model. We find all dilution factors to be within $1\sigma$ of zero, showing no indication of differential dilution in the photometric apertures.  We then fit the light curves with no dilution and found $R_p/R_* = 0.053^{+0.002}_{-0.001}$, which is within $1\sigma$ of the $R_p/R_*$ value found from the EXOFASTv2 global fit, confirming consistency with the TESS aperture crowding correction factor. We also find an orbital period of $P = 4.0337339^{+0.0000084}_{-0.0000077}$ days, which is within $2\sigma$ of the period we find from the joint model of the TESS plus radial velocity data in the global fit.

\subsection{Spectroscopic Observations} \label{sec:Spectroscopic Observations}
To measure the mass of the transiting object, we obtained high-resolution spectroscopy of TOI-1994 from ground-based telescopes to extract radial velocities (RVs). 

\begin{figure*}
    \centering
    \includegraphics[scale=.5]{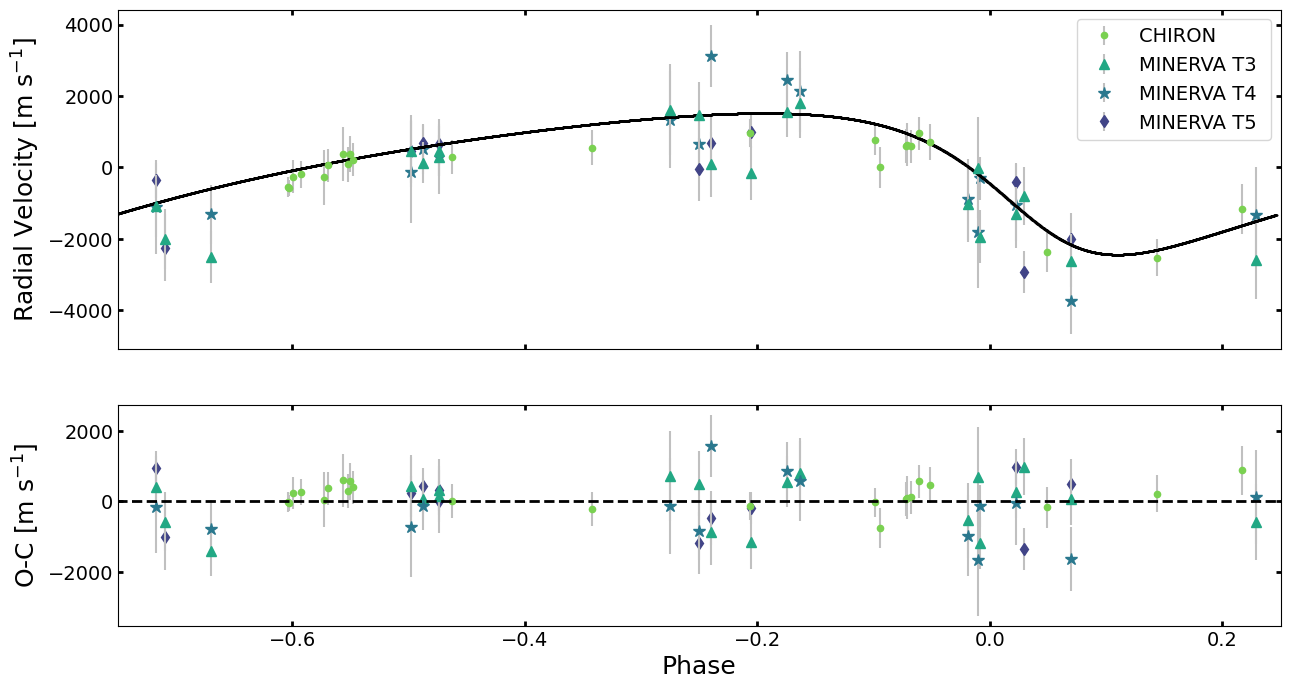}
    \caption{Radial velocities for TOI-1994 from CHIRON and three MINERVA-Australis telescopes. The black line represents the global fit model from the EXOFASTv2 described in Section \ref{sec:exofast}. The curve is phase-folded to a period of $4.0337142$ days. The residuals for the best fit model are shown below the radial velocity curve. }
    \label{fig:RV}
\end{figure*}

TOI-1994\,b was spectroscopically observed by the MINERVA-Australis telescope array in Toowoomba, Australia at the University of Southern Queensland’s Mount Kent Observatory. The telescopes T3, T4 and T5 were used for spectroscopic observations of TOI-1994. Each MINERVA-Australis telescope uses fiber optic cables to feed light into a spectrograph reaching resolution $R=80000$ for wavelengths from 480 to 630 nm \citep{2019PASP..131k5003A}. 

A total of 47 spectra were taken by MINERVA-Australis between 2021 June 5 and 2021 July 12 with an exposure time of 3600s. 


We also collected a 23 measurements with the CHIRON spectrograph \citep{Tokovinin:2013hia} on the 1.5-m SMARTS telescope at the Cerro Tololo Inter-American Observatory (CTIO) in Chile. The observations were obtained between 2021 November 6 and 2022 May 1, spanning the phase of TOI-1994\,b. We employed the fiber-fed image slicer to yield a resolving power of $R\sim80,000$\ between $4100$\ and $8700$\,\AA. The typical exposure time was 800 seconds and resulted in a mean S/N per resolution element of 65. We used the optimal extraction provided by the CHIRON team, described in \citet{Paredes:2021}, and extracted radial velocities by fitting line profiles derived by least-squares deconvolution of the observed spectra against synthetic templates \citep{Donati:1997, Zhou:2020}.

We note that TOI-1994 is fast-rotating, with $vsini>45$ km s$^{-1}$. This makes it difficult to extract reliable stellar parameters from the spectra, and affects the RV precision we are able to achieve.  In the global fit below, we adopt stellar parameters from the TIC for starting points of the analysis.

The first ten radial velocities from the MINERVA-Australis telescope array and CHIRON are listed in Table \ref{table:rv}. The full list of RV observations are available online as a machine readable table. The RVs with the best fit model are shown in Figure \ref{fig:RV} phase-folded using the period and epoch from the EXOFASTv2 global analysis described in section \ref{sec:exofast}.

\begin{deluxetable}{cccc}
\tabletypesize{\footnotesize}
\tablecaption{Radial velocity observations of TOI-1994 (all RV observations are available online as a machine readable table)}
\label{table:rv}
\tablewidth{0pt}
\tablehead{
\colhead{$\textrm{BJD}_{\textrm{{TDB}}} - 2450000.0$} & \colhead{RV (m s$^{-1}$)} & \colhead{$\sigma _{RV}$ (m s$^{-1}$)}& \colhead{Facility} }
\startdata
$9370.97041$ \dotfill & $-2600$  & $750$ & MINERVA T3 \\ 
$9370.97041$ \dotfill & $-2000$ & $710$ & MINERVA T5 \\
$9370.97041$ \dotfill & $-3740$  & $910$ & MINERVA T4 \\ 
$9377.92662$ \dotfill & $-150$  & $770$ & MINERVA T3 \\ 
$9377.92662$ \dotfill & $1000$  & $450$ & MINERVA T5 \\
$9524.84483$ \dotfill & $-1160$ & $700$ & CHIRON\\
$9533.84916$ \dotfill & $376$ & $520$ & CHIRON\\
$9533.86065$ \dotfill & $220$ & $466$ & CHIRON\\
$9535.86140$ \dotfill & $715$ & $510$ & CHIRON\\
$9537.85895$ \dotfill & $369$ & $755$ & CHIRON\\
\enddata 
\end{deluxetable}

\subsection{SOAR Speckle High Resolution Imaging} \label{SOAR}
The presence of unresolved luminous companions can complicate the characterization of a transiting system by diluting the true transit depth, or even creating a false positive signal \citep{Ciardi_2015}.  To search for any such companions, we used high resolution speckle imaging. 

\begin{figure}
    \centering
    \includegraphics[scale=.5]{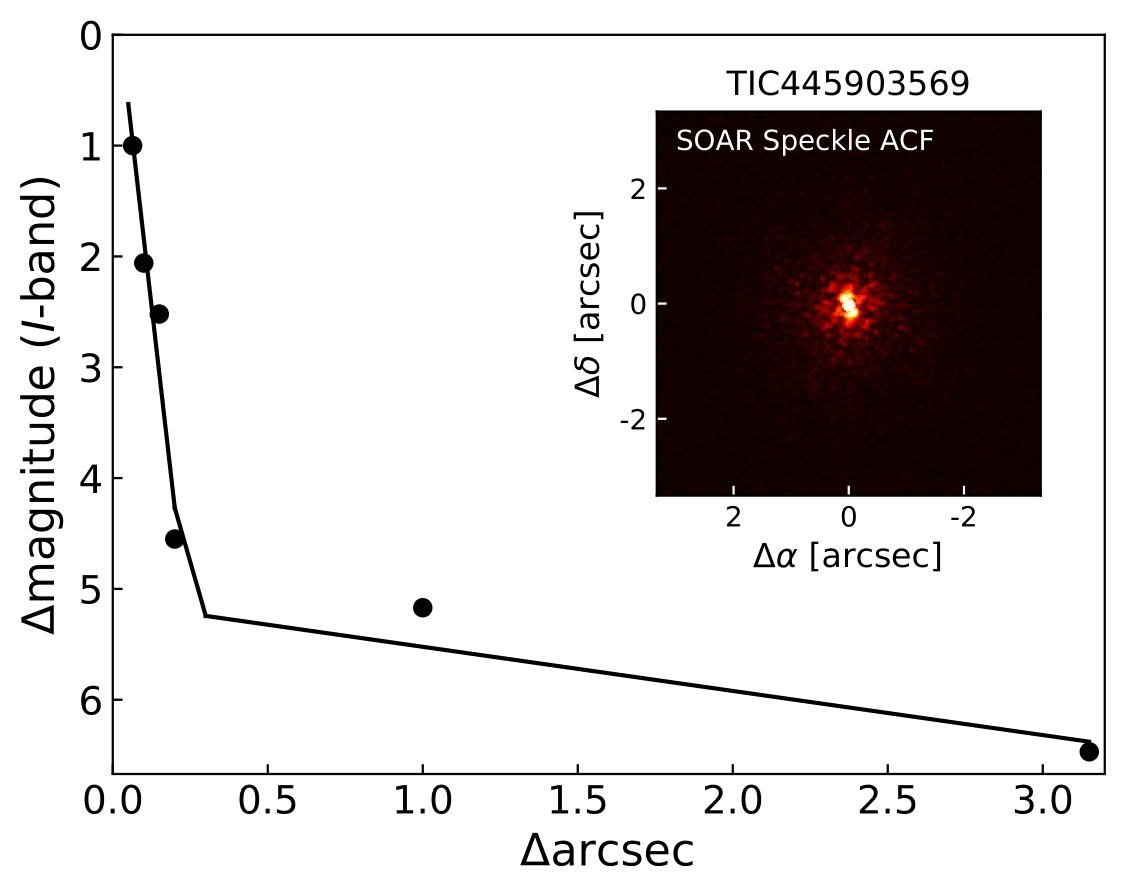}
    \caption{Speckle imaging from SOAR of TOI-1994. The points represent $5 \sigma$ detection sensitivity and the inset displays ACF from SOAR. No stellar companions are found within 3" down to the detection limits. }
    \label{fig:HRimaging}
\end{figure}

TOI-1994 was observed by the 4.1m Southern Astrophysical Research (SOAR) telescope using the \emph{I} filter on on October 31st, 2020. A description of the instrument is found in \citet{2018PASP..130c5002T} and observation details are described in \citet{Ziegler_2021}. Figure \ref{fig:HRimaging} displays the $5 \sigma$ detection sensitivity and inset auto correlation function (ACF) taken on October 31st, 2020 from SOAR. 

The speckle observations are sensitive to companion objects down to a $\Delta$ mag of 6.5 at 1". These observations show that there are no close stellar companions within 3" of TOI-1994. We therefore rule out the possibility of a false positive caused by a nearby stellar companion. 


\section{EXOFASTv2 Global fits} \label{sec:exofast}
We simultaneously modeled the transit photometry and RVs using EXOFASTv2, an exoplanet-fitting software package written in IDL \citep{2013PASP..125...83E, 2017ascl.soft10003E, https://doi.org/10.48550/arxiv.1907.09480}. This tool allows us to fit multiple data sets simultaneously using a Markov Chain Monte Carlo (MCMC) analysis.

EXOFASTv2 uses starting points for stellar properties, the spectral energy distribution, and MESA Isochrones and Stellar Tracks (MIST) in order to model the host star \citep{Dotter_2016, 2016ApJ...823..102C}. The spectral energy distribution we used as a part of the global model in EXOFASTv2 is given in Table \ref{table:SED}.  

We began the global modeling process by running an initial EXOFASTv2 fit using the \tess{} photometry and the RVs from the CHIRON spectrograph on the SMARTS 1.5-m telescope discussed in sections \ref{sec:TESS} and \ref{sec:Spectroscopic Observations}. This initial fit used starting points for stellar mass, radius and effective temperature from the TESS Input Catalog (TIC; \cite{2018AJ....156..102S, 2019AJ....158..138S}), and the period and epoch from TESS photometry.  We placed a wide Gaussian prior of $[Fe/H]=0.0\pm 1$ dex on metallicity, a Gaussian prior of $\varpi=1.94140 \pm 0.04211$ mas from Gaia DR2 on parallax, and an upper limit of $2.64275$ on \emph{V}-band extinction from \citet{1998ApJ...500..525S} and \citet{2011ApJ...737..103S}.

We ran another fit using EXOFASTv2 that incorporated everything from the initial fit, with the addition of RVs from the MINERVA-Australis telescope array, described in section \ref{sec:Spectroscopic Observations}. Each MINERVA-Australis telescope was fit as a separate spectrograph since the systematics are not consistent between them. The purpose of this second run was to fit for a slope over time in the RV data from both facilities. If a slope is detected, it could be a sign of a stellar companion with a much longer orbital period than TOI-1994\,b. This fit was consistent with a slope of zero over the 350 day baseline. Thus, the final fit presented here does not include a slope over time for the RV data.

The final fit, presented here, used \tess{} photometry, and RVs from the MINERVA-Australis telescope array and the CHIRON spectrograph on the SMARTS 1.5-m telescope. The description of this data is found in section \ref{sec:observations}. This fit used starting point for stellar properties, period, and epoch from the initial EXOFASTv2 fit. The same MIST evolution models and SED starting points described for the initial fit were used in this final fit. We used a smaller Gaussian prior on metallicity of $[Fe/H]=0.0\pm 0.5$ dex based on the results from the initial fit. We placed the same Gaussian prior on parallax from Gaia DR3, and the same upper limit on \emph{V}-band extinction described previously. The starting points and priors used in the final EXOFASTv2 fit are listed in Table  \ref{table:stellarparams}.

\begin{deluxetable*}{cccc}
\tablecaption{Properties From Liturature} 
\label{table:SED}
\tablewidth{0pt}
\tablehead{
\colhead{Identifiers: } }
\startdata
& TOI \dotfill  & TOI-1994 & \\
& TIC \dotfill & TIC 445903569 &\\
& TYCHO-2 \dotfill & TYC 8193-1368-1 & \\
& 2MASS \dotfill & J09484191-5145236 & \\
& HD \dotfill  & HD 298656 & \\
\hline
\hline
Parameter & Description & Value & Reference\\
\hline
$\alpha$ \dotfill & Right Ascension (RA) (J2000, epoch 2015.5) \dotfill  & 09:48:41.9 & \cite{GaiaDR3} \\
$\delta$ \dotfill & Declination (DEC) (J2000, epoch 2015.5)  \dotfill  & -51:45:23.63 & \cite{GaiaDR3}\\
T \dotfill & \tess{} mag \dotfill  & $10.2396 \pm 0.0061$  & \cite{2018AJ....156..102S, 2019AJ....158..138S}\\
G \dotfill & \emph{Gaia} G mag  \dotfill  & $10.4280 \pm 0.0200$  &  \cite{GaiaDR3}\\
 $\textrm{B}_\textrm{P}$ \dotfill  & \emph{Gaia} $\textrm{B}_\textrm{P}$ mag \dotfill   & $10.5787 \pm 0.0200$ & \cite{GaiaDR3}\\
$\textrm{R}_\textrm{P}$ \dotfill  & \emph{Gaia} $\textrm{R}_\textrm{P}$ mag \dotfill & $10.1965 \pm 0.0200$ & \cite{GaiaDR3}\\
$J$\dotfill  & 2MASS J mag \dotfill  & $9.915 \pm 0.024$ & \cite{2MASS} \\
 $H$ \dotfill & 2MASS H mag \dotfill & $9.843 \pm 0.023$ & \cite{2MASS}\\
$\textrm{K}_\textrm{S}$\dotfill  & 2MASS $\textrm{K}_\textrm{S}$ mag \dotfill  & $9.775 \pm 0.021$  & \cite{2MASS}\\
 $WISE1$\dotfill  &$WISE1$ mag \dotfill & $9.767 \pm 0.030$  & \cite{WISE}\\
 $WISE2$\dotfill  & $WISE$ mag \dotfill & $9.823 \pm 0.030$  & \cite{WISE}\\
 $WISE3$\dotfill  & $WISE3$ mag\dotfill  & $10.041 \pm 0.049$  & \cite{WISE}\\
$\mu_\alpha$ \dotfill & \emph{Gaia} DR3 proper motion in RA (mas yr$^-1$)\dotfill  & $-13.547 \pm  0.0577$
 & \cite{GaiaDR3}\\
$\mu_\delta$ \dotfill & \emph{Gaia} DR3 proper motion in DEC (mas yr$^-1$) & $-1.741 \pm 0.064
 $ & \cite{GaiaDR3}\\ 
$\varpi$ \dotfill & \emph{Gaia} DR3 Parallax (mas) \dotfill & $1.94140 \pm 0.04211$ & \cite{GaiaDR3}\\
\enddata
\end{deluxetable*}

Figure \ref{fig:TESS} displays the best-fit model for the \tess{} light curves in black, and Figure \ref{fig:RV} shows the best-fit radial velocity model in black. Table \ref{table:stellarparams} lists the derived stellar parameters and brown dwarf parameters from the final EXOFASTv2 global fit. 

From the EXOFASTv2 global fit analysis, we find that TOI-1994 is a hot, fast-rotating subgiant with a radius of $2.30^{+0.13}_{-0.12}$ $R_\odot\ $, mass of $1.86^{+0.18}_{-0.17}$ $M_\odot\ $, effective temperature of $7700^{+720}_{-410}$ K, and surface gravity of $3.982^{+0.067}_{-0.065}$\logg. 

The effective temperature and luminosity of TOI-1994 places it near the center of the instability strip for $\delta$ Scuti pulsations. Â Based on studies with Kepler (e.g., \cite{Bowman2018}, \cite{Murphy2019}), this implies a probability of around 50--70\% that TOI-1994 pulsates as a $\delta$ Scuti variable. We carried out Fourier analysis of the TESS light curve (using the residuals from the transit fit) on both the 2-min and 30-min data, but found no evidence for pulsations down to an amplitude limit of about 50 parts per million.


\begin{deluxetable*}{ccc}
\tablecaption{Median values and 68\% confidence intervals for the EXOFASTv2 global model.}
\label{table:stellarparams}
\tabletypesize{\footnotesize}
\tablehead{
\colhead{Symbol} & \colhead{Parameter} & \colhead{Value} }
\startdata
Priors:\\
$P$\dotfill & Period (days)\dotfill & $4.0337$  \\
$T_{C0}$\dotfill & Epoch ($\textrm{BJD}_{\textrm{{TDB}}}$)\dotfill & $2459290.01406$  \\
$\varpi$\dotfill & Parallax$^{*}$ (mas) \dotfill & $1.94140 \pm 0.04211$   \\
$[{\rm Fe/H}]$\dotfill & Metallicity$^{*}$ (dex)\dotfill & $0.0\pm .5$   \\
$A_V$\dotfill & V-band extinction$^{\dagger}$ (mag)\dotfill & $2.64275$   \\
\hline
Stellar Parameters:\\
$M_*$\dotfill & Mass ($M_\odot\ $)\dotfill & $1.86^{+0.18}_{-0.17}$\\
 $R_*$\dotfill & Radius ($R_\odot\ $)\dotfill & $2.30^{+0.13}_{-0.12}$\\
 $L_*$\dotfill & Luminosity ($L_\odot\ $)\dotfill & $16.7^{+6.3}_{-2.7}$\\
 $F_{Bol}$\dotfill & Bolometric Flux x 10$^{-9}$ (\fluxcgs)\dotfill &  $1.99^{+0.74}_{-0.31}$\\
 $\rho_*$\dotfill & Density (g cm$^{-3}$)\dotfill & $0.215^{+0.047}_{-0.038}$\\
 $\log{g}$\dotfill & Surface gravity (\logg)\dotfill & $3.982^{+0.067}_{-0.065}$\\
 $T_{\rm eff}$\dotfill & Effective Temperature (K)\dotfill & $7700^{+720}_{-410}$\\
 $[{\rm Fe/H}]$\dotfill & Metallicity (dex)\dotfill & $-0.01^{+0.30}_{-0.42}$\\
 $[{\rm Fe/H}]_{0}$\dotfill & Initial Metallicity$^{1}$ \dotfill & $0.07^{+0.28}_{-0.41}$\\
 $Age$\dotfill & Age (Gyr)\dotfill & $0.94^{+0.33}_{-0.31}$\\
 $EEP$\dotfill & Equal Evolutionary Phase $^{2}$ \dotfill & $361^{+26}_{-15}$\\
 $A_V$\dotfill & \emph{V}-band extinction (mag)\dotfill & $0.22^{+0.29}_{-0.16}$\\
 $\sigma_{SED}$\dotfill & SED photometry error scaling \dotfill & $2.67^{+1.2}_{-0.67}$\\
 $\varpi$\dotfill & Parallax (mas)\dotfill & $1.931\pm0.042$\\
 $d$\dotfill & Distance (pc)\dotfill & $517^{+12}_{-11}$\\
 \hline
~~~~Brown Dwarf Parameters:\\
 $P$\dotfill & Period (days)\dotfill & $4.0337142\pm0.0000051$\\
 $R_P$\dotfill & Radius ($R_J$)\dotfill & $1.220^{+0.082}_{-0.071}$\\
 $M_P$\dotfill & Mass ($M_J$)\dotfill & $22.1^{+2.6}_{-2.5}$\\
 $T_C$\dotfill & Time of conjunction ($\textrm{BJD}_{\textrm{{TDB}}}$)\dotfill & $2459290.01375^{+0.00079}_{-0.00084}$\\
 $T_0$\dotfill & Optimal conjunction Time$^{3}$ ($\textrm{BJD}_{\textrm{{TDB}}}$)\dotfill & $2458854.37263^{+0.00056}_{-0.00064}$\\
 $a$\dotfill & Semi-major axis (AU)\dotfill & $0.0613\pm0.0019$\\
 $i$\dotfill & Inclination (Degrees)\dotfill & $85.5^{+2.9}_{-2.5}$\\
 $e$\dotfill & Eccentricity \dotfill & $0.341^{+0.054}_{-0.059}$\\
 $\tau_{\rm circ}$\dotfill &Tidal circularization timescale (Gyr)\dotfill &$4.2^{+3.9}_{-2.1}$\\
 $\omega_*$\dotfill &Argument of Periastron (Degrees)\dotfill &$131.6^{+7.9}_{-6.7}$\\
 $T_{eq}$\dotfill &Equilibrium temperature (K)\dotfill &$2290^{+170}_{-110}$\\
 $K$\dotfill &RV semi-amplitude (m s$^{-1}$)\dotfill &$1970\pm170$\\
 $R_P/R_*$\dotfill &Radius of planet in stellar radii \dotfill &$0.05449^{+0.0012}_{-0.00084}$\\
 $a/R_*$\dotfill &Semi-major axis in stellar radii \dotfill &$5.72^{+0.39}_{-0.36}$\\
 $\delta$\dotfill &$\left(R_P/R_*\right)^2$ \dotfill &$0.002969^{+0.00013}_{-0.000091}$\\
 $\delta_{\rm TESS}$ \dotfill &Transit depth in TESS (fraction)\dotfill &$0.00328^{+0.00027}_{-0.00019}$\\
 $\tau$\dotfill &Ingress/egress transit duration (days)\dotfill &$0.00970^{+0.0020}_{-0.00095}$\\
 $T_{14}$\dotfill &Total transit duration (days)\dotfill &$0.1685^{+0.0021}_{-0.0019}$\\
 $b$\dotfill &Transit Impact parameter \dotfill &$0.32^{+0.17}_{-0.20}$\\
 $T_{S,14}$\dotfill &Total eclipse duration (days)\dotfill &$0.249^{+0.047}_{-0.049}$\\
 $\rho_P$\dotfill &Density (g cm$^{-3}$)\dotfill &$15.0^{+3.9}_{-3.2}$\\
 $logg_P$\dotfill &Surface gravity (\logg) \dotfill &$4.564^{+0.078}_{-0.080}$\\
 $T_S$\dotfill &Time of eclipse (\bjdtdb)\dotfill &$2459291.445^{+0.097}_{-0.10}$\\
 $e\cos{\omega_*}$\dotfill & \dotfill &$-0.222^{+0.036}_{-0.038}$\\
 $e\sin{\omega_*}$\dotfill & \dotfill &$0.255^{+0.056}_{-0.064}$\\
 $d/R_*$\dotfill &Separation at mid transit \dotfill &$4.02^{+0.61}_{-0.51}$\\
\enddata
\tablecomments{\begin{itemize} 
\item[$^*$] Indicates a Gaussian prior.
\item[$^\dagger$] Indicates an upper limit.
\item[$^1$] The Initial Metallicity is the metallicity of the star when it was formed.
\item[$^2$] The Equal Evolutionary Point corresponds to static points in a stars evolutionary history when using the MIST isochrones and can be a proxy for age. See §2
in \citet{Dotter_2016} for a more detailed description of EEP.
\item[$^3$] Optimal time of conjunction minimizes the covariance between $T_C$ and Period. This is the transit mid-point.
\end{itemize}}
\end{deluxetable*}

\section{Discussion} \label{sec:discussion}
\begin{figure}
    \centering
    \includegraphics[scale=.37]{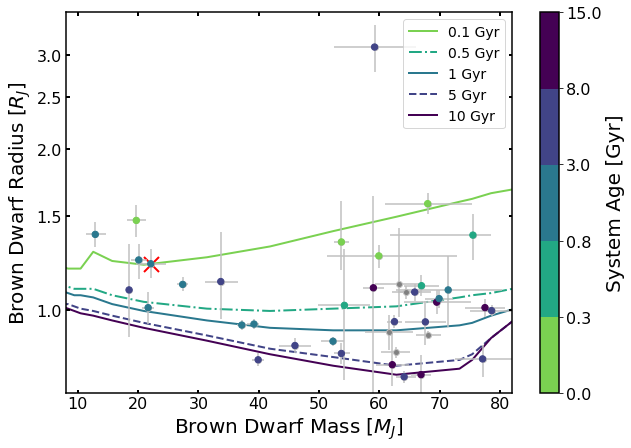}
    \caption{Mass-Radius diagram for 39 transiting brown dwarf companions. TOI-1994\,b is marked with a red 'x'. The colored lines represent isochrones for brown dwarfs at ages $0.1$, $0.5$, $1$, $5$ and $10$ Gyr \citep{2003A&A...402..701B}. Points are colored by the age of the system from publications. Grey represents systems published with a large age range.}
    \label{fig:iso}
\end{figure}

\begin{figure}
    \centering
    \includegraphics[scale=.4]{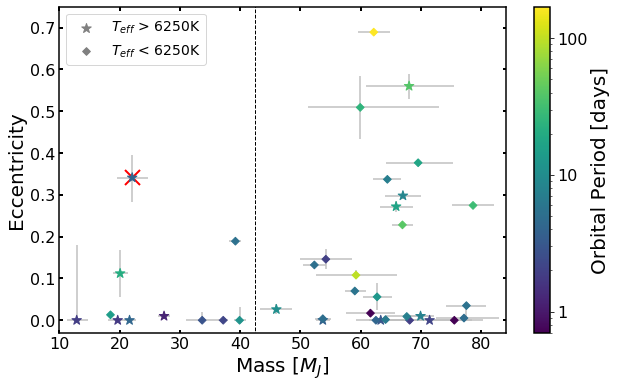}
    \caption{Eccentricity versus brown dwarf mass for 39 transiting brown dwarf companions. The color represents the orbital period of the system. Squares represent hosts with temperatures below the Kraft break ($T_{eff}\approx 6250$) and stars represent hosts with temperatures above the Kraft break. A vertical line shows the population split at 42.5 $M_J$ proposed by \citet{10.1093/mnras/stu134}. TOI-1994\,b is marked with a red 'x'. }
    \label{fig:ecc}
\end{figure}

\begin{figure}
    \centering
    \includegraphics[scale=.4]{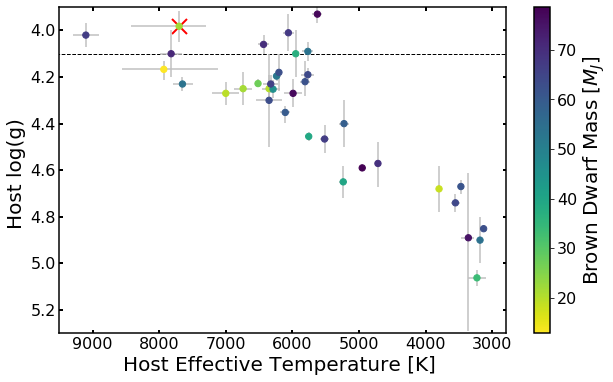}
   \caption{Surface gravity and effective temperature of 39 stars hosting transiting brown dwarf companions. The colors indicate the mass of the brown dwarf companions. TOI-1994\,b is marked with a red 'x'. The dashed horizontal dashed line indicates $log(g)=4.1$ as the transition between main sequence and evolved subgiant stars as in \citet{2018AJ....156..102S}. }
    \label{fig:logg}
\end{figure}

 We present the discovery of TOI-1994\,b, a low mass brown dwarf transiting an evolved host star with a moderately eccentric orbit.  We now compare TOI-1994\,b to 38 other transiting brown dwarf systems shown in Figures \ref{fig:iso}, \ref{fig:ecc}, and \ref{fig:logg}. We gathered properties for these other transiting brown dwarf systems from \citet{2022AJ....163...89C}, \citet{2022A&A...664A..94P},  \citet{2022MNRAS.514.4944C}, \citet{2022MNRAS.516..636S}, \citet{2023arXiv230109663V}, \citet{2022arXiv221013939L} and \citet{2021A&A...652A.127G} (see references therein). We define a brown dwarf as a substellar object between $13 M_J$ and $80M_J$ for this discussion.
   
Figure \ref{fig:iso} shows TOI-1994\,b on the mass-radius diagram with 38 other transiting brown dwarfs with masses between 13 and 80 $M_J$. We also show isochrones for 
brown dwarfs that have cooled due to the lack of hydrogen fusion with ages $0.1$, $0.5$, $1$, $5$ and $10$ Gyr \citep{2003A&A...402..701B}. Age affects the radius of a brown dwarf depending on its mass, with older objects having smaller radii. We note that there is further work being done in order to refine radii of transiting brown dwarfs, such as \citet{2023MNRAS.519.5177C}. 

We determine a tight age constraint of $0.94^{+0.33}_{-0.31}$ Gyr using MIST models in the EXOFASTv2 fit, since TOI-1994 is a subgiant star.  We expect that the brown dwarf companion formed at the same time as its stellar host and also has an age of $0.94$ Gyr. Using the tight age constraint, along with precise mass, radius and eccentricity, we can directly test substellar evolution models. In Figure \ref{fig:iso}, TOI-1994\,b falls along a younger isochrone of 0.1 Gyr. This indicates that the radius of TOI-1994\,b is larger than the models predict for a $0.94$ Gyr brown dwarf. The larger radius of TOI-1994\,b may be caused by an increase in luminosity of the host star causing re-inflation as it evolves off the main sequence. This is a method of re-inflation proposed for hot giant planets as in \citet{Thorngren_2021} and \citet{Lopez_2016}. However, planet mass may effect re-inflation, with less massive objects experiencing more dramatic inflation \citep{Thorngren2018, Thorngren_2021}. Otherwise, as pointed out in \citet{2001ApJ...548..466B}, tidal heating of short period planets in eccentric orbits can cause inflated radii.  It is possible that the inflated radius of TOI-1994\,b can be explained by its orbital dynamics. More evidence would be required to fully confirm whether these planet inflation mechanisms could extend to include brown dwarfs. 

In addition to radius and mass, period and eccentricity also give insight to the dynamical nature of the transiting system. The orbital eccentricity for close-in planets and stars with orbital periods less than $\sim 10$ days may become circularized from interactions with their host stars over longer timescales \citep{1981A&A....99..126H, Adams_2006, 1966Icar....5..375G}. 
 
With an age of $0.94$ Gyr and period of $4.0337142$ days, we explore whether the system should be circularized. In order to estimate the circularization timescale, $\tau_{circ}$, for TOI-1994\,b, we used Equation 3 from \citet{Adams_2006}: 

\begin{multline}
\tau_{circ} \approx 1.6 \textrm{ [Gyr] } \left(\frac{Q_p}{10^6}\right) \left(\frac{M_P}{M_{J}}\right) 
\\ \quad \times \left(\frac{M_*}{M_\odot}\right)^{-3/2} \left(\frac{R_P}{R_{J}}\right)^{-5} \left(\frac{a}{0.05 \textrm{ [AU]}}\right)^{13/2}
    \label{eq:circ}
\end{multline}

Using the EXOFASTv2 values from Table \ref{table:stellarparams}, and an estimate of $\frac{Q_P}{10^6} \approx 1$ where $Q_P$ is the tidal quality factor, we found that the circularization timescale for TOI-1994\,b is $4.2$ Gyr, with large uncertainties of $+3.9$ and $-2.1$ Gyr. However, even with the large uncertainty, the 2 sigma lower bound is still greater than the age of TOI-1994\,b of $0.94^{+0.33}_{-0.31}$ Gyr found using global analysis. Thus, we would not expect the system to be circularized. This is confirmed by the eccentricity of $0.314^{+0.054}_{-0.059}$. TOI-1994\,b is moderately eccentric compared to other brown dwarfs with similar mass and orbital period. Figure \ref{fig:ecc} displays the relationship between brown dwarf mass, eccentricity and period for 39 transiting brown dwarf companions including TOI-1994\,b. The brown dwarfs with orbital periods shorter than 10 days are mostly circularized as expected, with TOI-1994\,b as an exception.

 \citet{10.1093/mnras/stu134} suggests that brown dwarfs can be split into two different populations at 42.5 $M_J$ based on their eccentricity distribution. They found that brown dwarfs with lower masses match the eccentricity distribution for planets, and the higher mass match the eccentricity distribution for stars. For masses below 42.5 $M_J$, the more massive a brown dwarf is, the lower eccentricity it tends to have. The brown dwarf eccentricity distribution is also similar to giant planet eccentricity distributions in more recent papers, such as in \citet{2023MNRAS.tmp..609R}. Above 42.5 $M_J$, the eccentricities of brown dwarf companions are more widely distributed. \citet{10.1093/mnras/stx334}, \citet{refId0} and \citet{2021A&A...652A.127G} conducted a similar analysis using more brown dwarfs and supported the results of \citet{10.1093/mnras/stu134}. This split in population may arise from different formation mechanisms, with lower mass brown dwarfs forming like planets, and higher mass brown dwarfs forming like stars. Figure \ref{fig:ecc} displays the relationship between brown dwarf mass, eccentricity and period for the brown dwarf population, and places a line at 42.5 $M_J$ to show the split in population proposed by \citet{10.1093/mnras/stu134}. The split in population can be clearly seen, with higher mass brown dwarfs being more distributed in the parameter space. 
 
TOI-1994\,b has a mass lower than $42.5$ $M_J$ and therefore should follow the eccentricity distribution for planets, but it is more eccentric than expected. With the addition of TOI-1994\,b, it is clear that more transiting brown dwarfs with precise masses are required to verify or refute the population split at $42.5$ $M_J$ determined from eccentricity distributions.

 
 The higher eccentricity of TOI-1994\,b could be potentially explained by gravitational interactions with a stellar companion. We searched for signs of this in the speckle imaging, and found no stellar companions within 3". We also checked for a slope in the RV data over time, and found that the fit is consistent with a slope of zero over the year of data presented here. Further analysis on a longer time baseline would be necessary to detect the presence of any widely separated stellar companion.

Figure \ref{fig:logg} shows the effective temperature and surface gravity of the 39 host stars colored by the mass of their transiting brown dwarf companions. A horizontal line is placed at $log(g)=4.1$ to indicate the transition between main sequence and evolved hosts as defined by \citet{2018AJ....156..102S}. There are only four host stars in this set with $log(g)<4.1$ including TOI-1994.

\citet{2018ApJ...861L...5G} suggests that close-in giant planets orbiting evolved stars have more eccentric orbits because of tidal interactions with the evolved host. The hot giant planets may go through an eccentric phase when their orbits shrink faster than they circularize. If this is extended to include low mass brown dwarfs orbiting evolved stars, it would mean that close-in brown dwarfs orbiting evolved stars also have higher eccentricity. There are currently only four evolved hosts for brown dwarfs. Two of these brown dwarfs, including TOI-1994, have moderately eccentric orbits that support this hypothesis. Figure \ref{fig:logg} shows the host surface gravity versus host effective temperature, with colors representing the mass of the brown dwarf companion. The figure displays that there are only four evolved hosts, using the TESS definition. TOI-1994\,b has the lowest mass out of the evolved hosts. It is clear that more brown dwarfs orbiting evolved stars need to be discovered before any real conclusions can be made about how the host evolution effects the brown dwarf in the systems.


\acknowledgements
This work was funded by NASA XRP grant 80NSSC18K0544. Funding for the TESS mission is provided by NASA's Science Mission Directorate. We acknowledge the use of public TESS data from pipelines at the TESS Science Office and at the TESS Science Processing Operations Center. Resources supporting this work were provided by the NASA High-End Computing (HEC) Program through the NASA Advanced Supercomputing (NAS) Division at Ames Research Center for the production of the SPOC data products. This paper includes data collected by the TESS mission that are publicly available from the Mikulski Archive for Space Telescopes (MAST).

This research has made use of the SIMBAD database, operated at CDS, Strasbourg, France. This research has made use of NASA's Astrophysics Data System Bibliographic Services. This research has made use of the NASA Exoplanet Archive, which is operated by the California Institute of Technology, under contract with the National Aeronautics and Space Administration under the Exoplanet Exploration Program. This research has made use of the Exoplanet Follow-up Observation Program website, which is operated by the California Institute of Technology, under contract with the National Aeronautics and Space Administration under the Exoplanet Exploration Program.

MINERVA-Australis is supported by Australian Research Council LIEF Grant LE160100001, Discovery Grants DP180100972 and DP220100365, Mount Cuba Astronomical Foundation, and institutional partners University of Southern Queensland, UNSW Sydney, MIT, Nanjing University, George Mason University, University of Louisville, University of California Riverside, University of Florida, and The University of Texas at Austin.

We respectfully acknowledge the traditional custodians of all lands throughout Australia, and recognise their continued cultural and spiritual connection to the land, waterways, cosmos, and community. We pay our deepest respects to all Elders, ancestors and descendants of the Giabal, Jarowair, and Kambuwal nations, upon whose lands the Minerva-Australis facility at Mt Kent is situated.
GZ is supported by Australian Research Council Discovery Early Career Researcher Award DE210101893.

\facilities{\tess, MINERVA-Australis, CTIO, SOAR, Exoplanet Archive}
\software{Astropy, lightkurve, EXOFASTv2}

\newpage

\bibliographystyle{aasjournal}
\bibliography{bibliography}

\end{document}